\documentclass[preprint, amsmath,amssymb,aps,keywords
prb,
]{revtex4-2}

\usepackage{graphicx,color}
\usepackage{graphicx}
\usepackage{dcolumn}
\usepackage{bm}
\usepackage{hyperref}
\usepackage[utf8]{inputenc}
\usepackage[T1]{fontenc}
\usepackage{amsmath,amssymb,graphicx}

\begin{document}


\title{Dirac fermions on wires confined to the graphene M\"{o}bius strip}

\author{L. N. Monteiro}
 \affiliation{Universidade Federal do Cear\'{a} (UFC), Departamento de F\'{i}sica, Campus do Pici, Fortaleza-CE, 60455-760, Brazil.}
\author{J. E. G. Silva}%
\affiliation{Universidade Federal do Cariri (UFCA), Cidade Universit\'{a}ria, Juazeiro do Norte - CE, 63048-080, Brazil.}
\author{C. A. S. Almeida}%
\affiliation{Universidade Federal do Cear\'{a} (UFC), Departamento de F\'{i}sica, Campus do Pici, Fortaleza-CE, 60455-760, Brazil.}





\date{\today}

\begin{abstract}
We investigate the effects of the curved geometry on a massless relativistic electron constrained to a graphene strip with a M\"{o}bius strip shape. The anisotropic and parity violating geometry of the M\"{o}bius band produces a geometric potential that inherits these features. By considering wires along the strip width and the strip length, we find exact solutions for the Dirac equation and the effects of the geometric potential on the electron were explored. In both cases, the geometric potential yields to a geometric phase on the wave function. Along the strip width, the density of states depends on the direction chosen for the wire, a consequence of the lack of axial symmetry. Moreover, the breaking of the parity symmetry enables the electronic states to be concentrated on the inner or on the outer portion of the strip. For wires along the strip length, the nontrivial topology influences the eigenfunctions by modifying their periodicity. It turns out that the ground state has period of $4\pi$ whereas the first excited state is a $2\pi$ periodic function. Moreover, we found that the energy levels are half-integer multiples of the energy of the ground state. 
\end{abstract}

\maketitle


\section{Introduction}

Since its discover, graphene has startled researchers due to its outstanding electronic, mechanical and thermal properties \cite{graphene,Geim}. This single layer sheet of carbon exhibits high carrier mobility (vanishing effective electron mass) \cite{Novoselov2004} and thus, the electron is described as a massless chiral Dirac fermion on a flat surface \cite{electronic}. As a result, graphene offers a bridge between condensed matter physics and quantum field theory in two dimensions \cite{katsnelson}. Indeed, graphene produces well-known relativistic effects, such as the zitterbewegun \cite{zitter} and the Klein paradox \cite{klein}. More recently, relativistic effects were found in Weyl semimetals \cite{weyl}, Majorana fermions \cite{majorana}, Bogoliubov particles \cite{bogoliubov} and Kagome crystals \cite{kagome}.

By bending the two dimensional sheet or considering the strain effects on the membrane, graphene also becomes a table-top laboratory for curved spaces phenomena \cite{graphenegeometry,diracgraphene}. In fact, the curved geometry or the strain tensor modify the effective Hamiltonian leading to a position-dependent Fermi velocity \cite{fermivelocity}. Moreover, the coupling between the Dirac fermion and the curved geometry/strain provides pseudo gauge fields whose effects depend on the particular geometry/strain \cite{gaugeingraphene,strain,pseudogaugefields,Diracmagnons}. Despite being two dimensional, graphene naturally displays ripples\cite{ripples,wrinkles} and corrugations \cite{corrugated}. These deformations of the surface significantly modify the electronic and thermal graphene properties \cite{ripple2}. The effects of the curved geometry on the electronic properties, the so called \textit{curvatronics} has been explored in different geometries, such as the cone \cite{cone}, helicoid strip \cite{watanabe,helix}, the catenoid bridge \cite{catenoid,catenoid2} and the torus \cite{torus}.

Another noteworthy effect produced by the curved geometry is the so-called geometric phase \cite{geometricphase}. The edge states \cite{geometricphase2} or the presence of defect, such as disclinations \cite{geometricphase3}, produce geometric phases modifying the electronic properties. On conical graphene surfaces, a constant geometric gauge field yields to a geometric phase depending on the conical deficit angle \cite{graphenecone}. Furthermore, the out-of-plane deformations of the graphene layer also lead to modifications on the optical conductivity due to Aharonov-Bohm type interference \cite{ABoptics}.

Besides the curved geometry, nontrivial topology also plays a central role on the electronic features of the low-dimensional system \cite{mobiusmolecues}. For instance, performing a $\pi$ twist on one end of a graphene ribbon and connecting one end to the other we obtain a graphene M\"{o}bius strip \cite{spivak,mobiuselasticity}. This graphene membrane behaves as a topological insulator with stable edge states \cite{mobiustopologicalinsulator}. Moreover, the half twist modifies the rotational invariance of the fields and particles constrained to move along the M\"{o}bius band \cite{mobiussymmetry}. Several investigations on the formation \cite{mobiusstrip1}, stability \cite{mobiusstrip2}, charge transfer \cite{mobiusstrip3}, magnetic properties \cite{mobiusstrip4}, quantum spin-Hall effect \cite{mobiusstrip5,hans} of the graphene M\"{o}bius strip were performed. These properties are rather different from those obtained from usual cylindrical graphene rings \cite{mobiusring}. Indeed, the curvature of the M\"{o}bius strip modifies the energy spectrum of non-relativistic electrons, whose isotropic and parity symmetries are broken \cite{mobiusschrodinger}. For relativistic electrons, the influence of the a flat M\"{o}bius strip was investigate by assuming nontrivial boundary conditions \cite{mobiusdirac}.

In this work we study the effects of the graphene M\"{o}bius strip curved geometry on effective massless relativistic electrons. Unlike some previous works that used discrete tight-bind approach \cite{mobiustopologicalinsulator,mobiusstrip1,mobiusstrip2}, we employed a continuum analysis wherein an effective massless Dirac fermion is constrained to a curved surface. Since Dirac equation on a curved surface does not couple with the curvature \cite{burgess,Jensen}, the effects of the curved geometry steam from the spinorial connection which acts as a pseudo-magnetic potential induced by the curved geometry \cite{vozmediano}. As a result, a geometric phase determines the density of states for wires along the M\"{o}bius strip width. The breaking of the parity and isotropy symmetries yields to states localized on the inner or outer portion of the strip depending on the angle chosen for the wire. For wires along the strip length (angular direction), the geometric phase has an Aharonov-Bohm like effect, not modifying the density of states for a single electron. Nevertheless, the ground state wave function has a period of $4\pi$, whereas the first excited state is a $2\pi$ periodic function, due to the strip twist. Furthermore, the nontrivial topology also leads to an energy spectrum given as half-integer multiples of the ground state energy.
These results agree with the modified boundary conditions analysis performed in Ref.(\cite{mobiusschrodinger,mobiusdirac}).

This work is organized as the following. In the section (\ref{sec2}) we briefly review the main definitions and properties of the M\"{o}bius strip geometry, such as the metric, curvatures and connection. In the section (\ref{sec3}),
we obtain the effective Hamiltonian for the massless Dirac particle on the surface and derive the expression for the geometric potential which depends on the geometric connection. In the section (\ref{sec4}) we obtain the exact solutions for the electron on wires along the width and the length of the strip. The energy levels and the effects of the nontrivial topology on the states and spectrum is discussed. Finally, additional comments and perspectives are outlined in section (\ref{sec5}).

\section{M\"{o}bius strip and its properties}
\label{sec2}

In this section we present the coordinate system to describe the graphene M\"{o}bius strip and study its main geometric features, such as the gaussian and the mean curvatures and the connection one-forms.

\begin{figure}[ht!]
    \centering
    \includegraphics[scale=0.6]{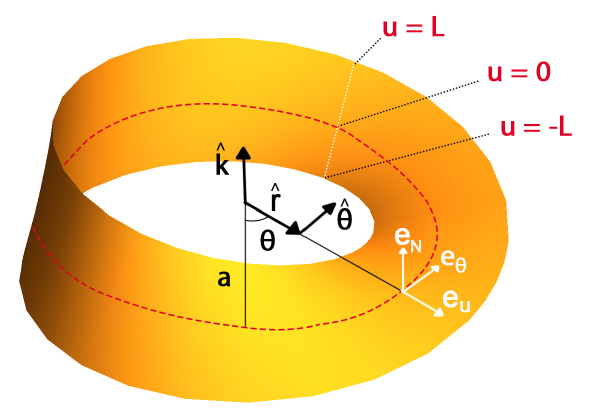}
    \caption{Coordinates and base vectors on the Möbius strip.}
    \label{fig:strip}
\end{figure}

The graphene M\"{o}bius strip is a surface constructed by joining the two ends of a graphene ribbon after twist one end by a $\pi$ rotation, as shown in the fig.(\ref{fig:strip}) \cite{spivak}. This surface can be described by the following coordinate system \cite{spivak,mobiusschrodinger}
\begin{align}\nonumber
\mathbf{r}(u,\theta) &=\left(a+u\cos\frac{\theta}{2}\right)\cos \theta \, \hat{\mathbf{i}}+\left(a+u\cos\frac{\theta}{2}\right)\sin \theta\hat{\mathbf{j}}+u\sin\frac{\theta}{2}\hat{\mathbf{k}}  \\
&=\left(a+u\cos\frac{\theta}{2}\right)\hat{\mathbf{r}}+u\sin\frac{\theta}{2}\hat{\mathbf{k}},
\label{parametrization}
\end{align}
where $2L$ is the strip width , $-L\leq u\leq L$ is a coordinate along the width, and $0\leq\theta\leq2\pi$ is the angular coordinate. The coordinate $u$ is the distance between the center of the strip and its endpoint, as shown in the Fig. \ref{fig:strip}. Furthermore, the parameter $a$ is the radius of the central circle and the radial unit vector is $\hat{\mathbf{r}}=\cos\theta\hat{\mathbf{i}}+\sin\theta \hat{\mathbf{j}}$.
A remarkable feature of the M\"{o}bius strip is its non-orientability. Indeed, performing a $2\pi$ rotation on the outer edge of the strip ($u=L$), the outer edge becomes the inner edge $(u=-L)$. Throughout the work, we use $L=1$.

\begin{figure}[htb]
\begin{center}
\includegraphics[scale=0.24]{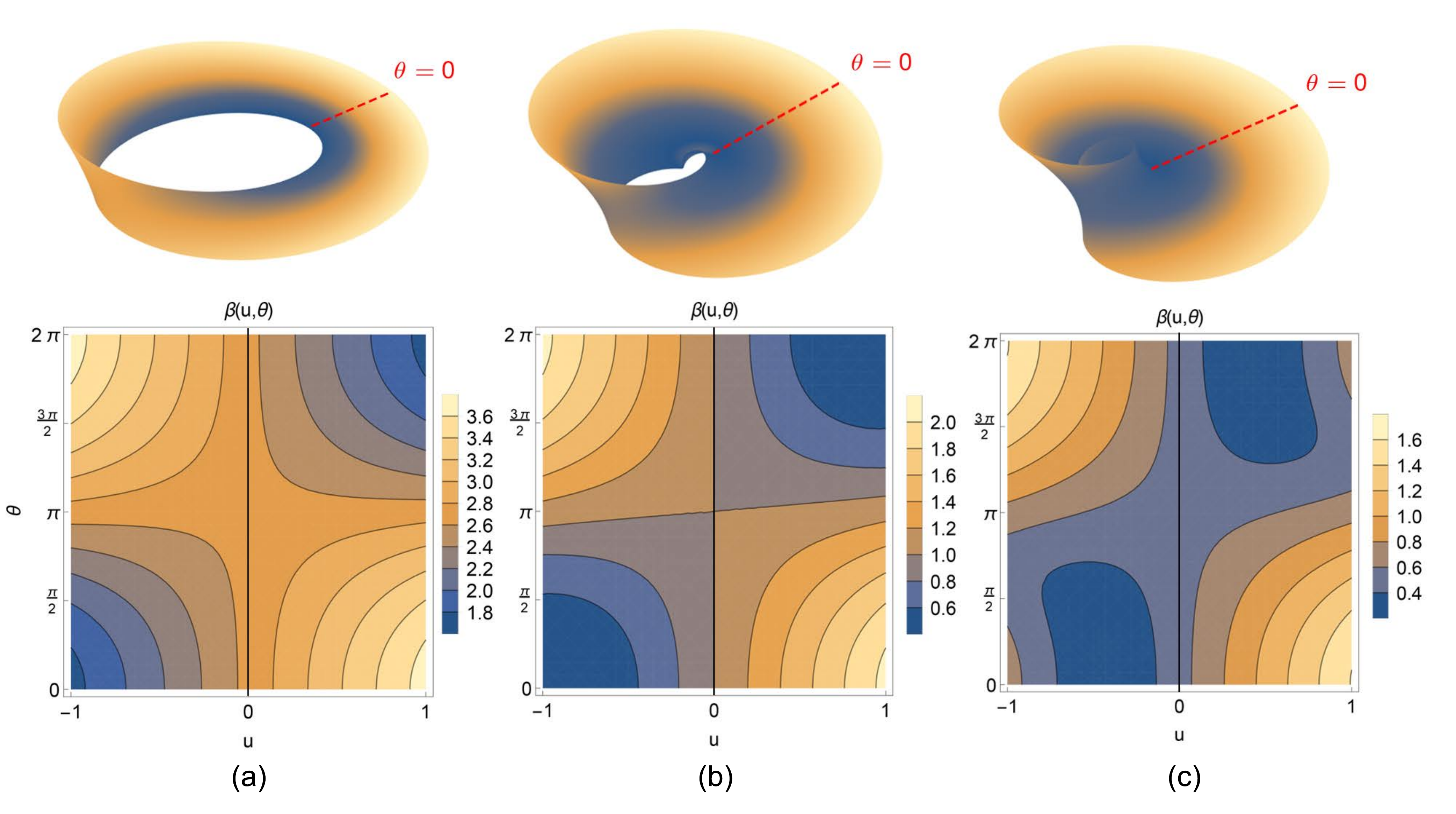}
\end{center}
\caption{Angular metric function $\beta(u,\theta)$ for $L=1$, with L/a = 0.375 (a), L/a = 1 (b) and L/a = 1.89 (c), which encodes the geometrical properties of the M\"{o}bius graphene strip. The function shows a twisted parity symmetry.}
\label{betagraphic}
\end{figure}

The tangent vectors, defined as  $ e_i = \frac{\partial \mathbf{r}}{\partial x^i}$, have the form
\begin{eqnarray}
e_u & = & \frac{\partial\textbf{r}}{\partial u}=\cos\frac{\theta}{2}\hat{r}+\sin\frac{\theta}{2}\hat{k}\\
e_\theta & = & \frac{\partial\textbf{r}}{\partial\theta}=\bigg(a+u\cos\frac{\theta}{2}\bigg)\hat{\theta}-\frac{u}{2}\sin\frac{\theta}{2}\hat{r}+\frac{u}{2}\cos\frac{\theta}{2}\hat{k}.
\end{eqnarray}
From the tangent vectors, we can define the 
M\"{o}bius strip metric $g_{ij}$ as
\begin{align}
g_{ij}=e_i\cdot e_j=\begin{pmatrix}
1 & 0\\ 0 & \beta^2(u,\theta), \end{pmatrix}
\end{align}
where the angular metric factor $\beta$ is given by \cite{mobiusschrodinger}
\begin{equation}
\label{beta}
\beta(u,\theta)=\sqrt{\frac{u^2}{4}+\left(a^2+u\cos\left(\frac{\theta}{2}\right) \right)^2}.
\end{equation}
In the fig.(\ref{betagraphic}) we plot the angular metric function $\beta(u,\theta)$ for some values of the inner radius $a$ with $L$ fixed to $L=1$. In fig.(\ref{betagraphic})(a) we choose $a$ such that $L/a=0.375$, i.e., $a=2.66$, for which nanorings were obtained \cite{ABring}. As we increase the ratio $L/a$, as for $L/a=1$ in fig. (\ref{betagraphic})(b) and for $L/a=1.89$ in fig. (\ref{betagraphic})(c) the M\"{o}bius strip becomes more compact. The ratio $L/a=1.89$ is a critical value for a M\"{o}bius strip due to mechanical properties \cite{mobiusstrip1}.


Thus, the $2+1$ infinitesimal line element has the form
\begin{equation}
    ds^2 = -dt^2 + du^2 + \beta^2 (u,\theta) d\theta^2.
    \label{mobiusmetric}
\end{equation}
It is worthwhile to mention that the metric tensor for a cylinder (nanotubes) has the form shown in  Eq.(\ref{mobiusmetric}) with $\beta=a$, whereas for the conical surface (graphitic cone), $\beta=\alpha u$, where $\alpha$ is the so-called angular deficit \cite{cone}. Other graphene based surface, such as the helicoid \cite{watanabe}, catenoid \cite{catenoid,catenoid2} and the torus \cite{torus} can also be described by the metric in Eq.(\ref{mobiusmetric}).

The anisotropy of the M\"{o}bius strip breaks some surfaces symmetries, such as the parity symmetry, as we can see by $\beta(-u,-\theta)\neq \beta(u,\theta)$. However, a sort of modified M\"{o}bius parity symmetry holds, where 
\begin{equation}
\label{mobiusparity}
\beta(-u,2\pi-\theta)= \beta(u,\theta).
\end{equation}
This modified parity symmetry steams from the twist performed on the strip. In the fig.(\ref{betagraphic}), we plotted the angular metric component $\beta$ for $a=1$ and $-1\leq u \leq 1$, where the M\"{o}bius parity symmetry in Eq.(\ref{mobiusparity}) is shown.

\begin{figure}[!htb]
\begin{center}
\includegraphics[scale=.24]{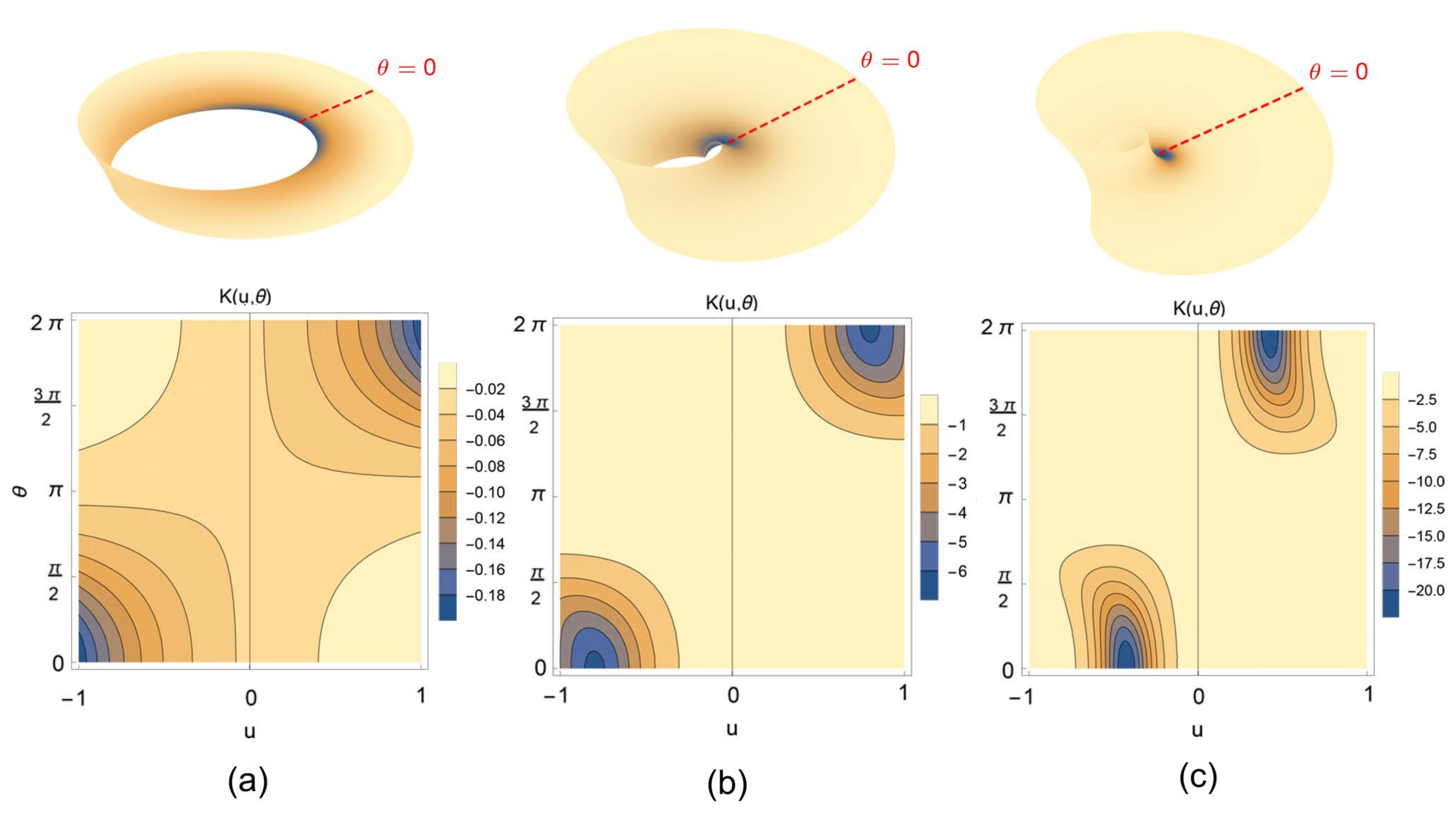}
\end{center}
\caption{Gaussian curvature on the M\"{o}bius strip, with L/a = 0.375 (a), L/a = 1 (b) and L/a = 1.89 (c), showing how the curvature is intrinsically distributed over the surface.} 
\label{gausscurvature}
\end{figure}

\begin{figure}[!htb]
\begin{center}
\includegraphics[scale=.24]{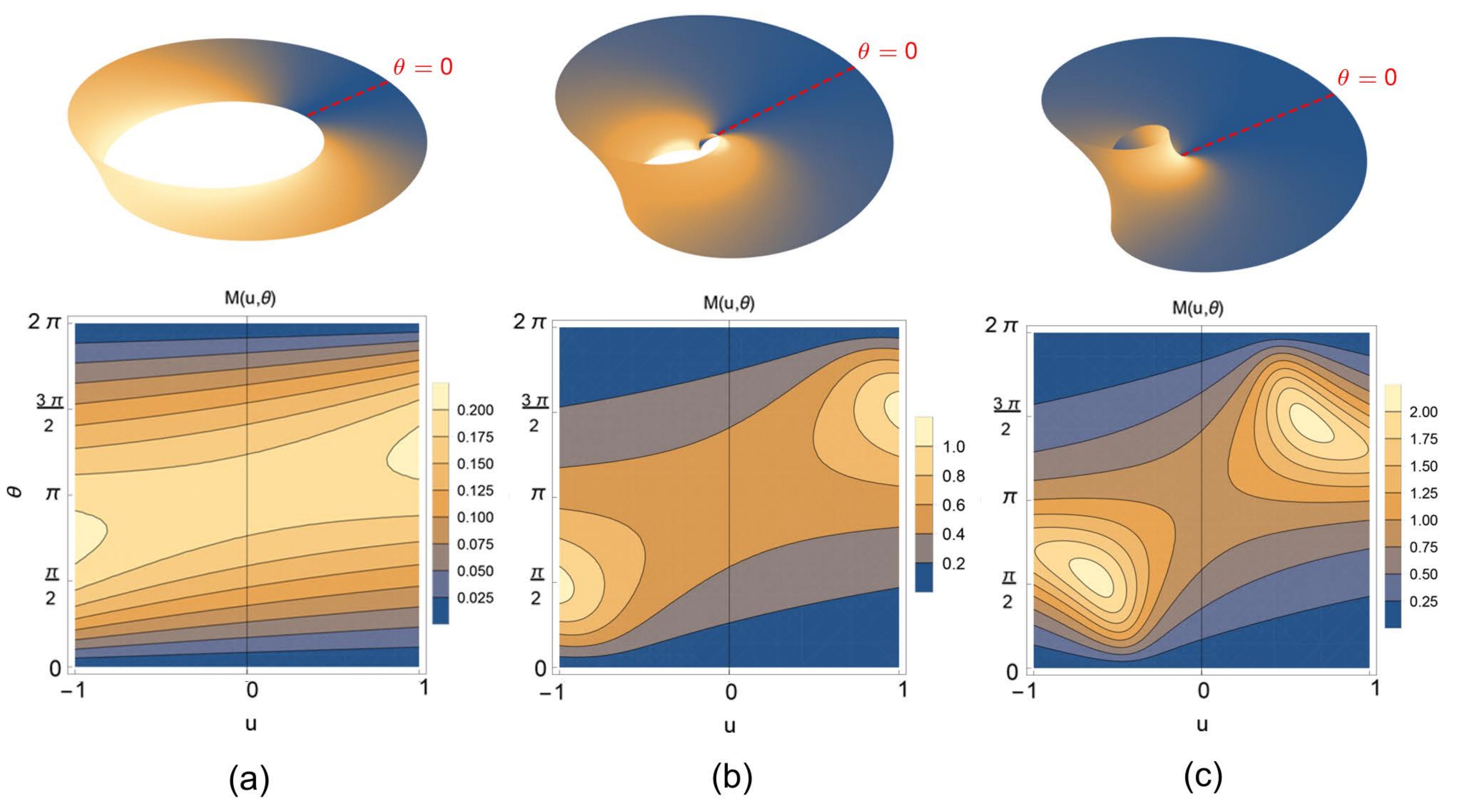}
\end{center}
\caption{Mean curvature on the M\"{o}bius strip with L/a = 0.375 (a), L/a = 1 (b) and L/a = 1.89 (c), showing how the curvature is extrinsically distributed over the surface.}
\label{meancurvature}
\end{figure}

The M\"{o}bius strip curvatures also exhibit the twisted parity symmetry in Eq.(\ref{mobiusparity}). Indeed, the expressions for the mean curvature $M$ and the gaussian curvature $K$ are given by \cite{mobiusschrodinger}
\begin{align}
    M=\frac{2(2(a^2+u^2)+4au \cos\frac{\theta}{2}+u^2 \cos\theta)\sin\frac{\theta}{2}}{[4a^2+3u^2+2u(4a\cos\frac{\theta}{2}+u\cos\theta)]^\frac{3}{2}},
\end{align}
and
\begin{align}
    K=-\frac{1}{\beta^2}\partial^2_u \beta=-\frac{4 a^2}{(4a^2+u^2+8au\cos(\theta/2)+4u^2\cos^2(\theta/2))^2}.
\end{align}
The behaviour of the gaussian and the mean curvatures are shown in the fig.(\ref{gausscurvature}) and in the fig.(\ref{meancurvature}), respectively. We adopt the same values for the ratio $L/a$ as done for the $\beta$ function, i.e., $L/a=0.375$ (a), $L/a=1$ (b) and $L/a=1.89$ (c).  Note that the curvatures profiles depend on the ratio $L/a$. Unlike the cylindrical nanorings, where the gaussian curvature vanishes and the mean curvature is constant, the distribution of the curvatures on the M\"{o}bius strip depends on the position $u,\theta$ and the ration $L/a$.

As we will see in the next section, the massless relativistic electron couples with the surface connection instead of the curvature. Accordingly, in order to define the fermion dynamics on the surface, we have to adopt the so-called \textit{vielbein} formalism \cite{diracgraphene}. Indeed, consider a set of matrices $e^{a}_\mu$ such that \cite{cone}
\begin{equation}
g_{\mu\nu}=e^{a}_\mu e^{b}_\nu \eta_{ab}.
\end{equation}
In $(2+1)-D$, the matrices $e^{a}_\mu$ are $SO(1,2)$ invariant and they are called \textit{dreinbeins}. Using the \textit{dreinbeins} we can define a local moving co-frame as $e^{a}=e^{a}_\mu dx^\mu$. For the M\"{o}bius metric in Eq.(\ref{mobiusmetric}), the moving co-frame has the form
\begin{eqnarray}
e^{0} & = & dt\nonumber\\
e^{1} & = & \cos\theta du -\beta \sin\theta d\theta\nonumber\\
e^{2} & = & \sin\theta du + \beta \cos\theta d\theta.
\label{dreinbeins} 
\end{eqnarray}
The presence of the trigonometric functions in eq.(\ref{dreinbeins}) reveals the local rotational invariance on the surface. Using the torsion-free condition, $de^{a}+\omega^{a}_{b}\wedge e^{b}=0$, the only non-vanishing connection 1-form $\omega^{a}_{b}=\Gamma^{a}_{\phantom{d}b\mu}dx^{\mu}$ is given by
\begin{equation}
\label{connection}
\omega^{1}_{2}=-(\partial_u \beta -1)d\theta.
\end{equation} 
Note that for a conical graphitic surface, $\beta=\alpha u$, and then, $\omega^{1}_{2}=-(\alpha -1)d\theta$ \cite{cone}.


\section{Fermions coupled to the M\"{o}bius strip}
\label{sec3}

Once we reviewed the main geometric features of the graphene M\"{o}bius strip, in this section we describe  how the effective massless electron couples to the graphene M\"{o}bius strip. We study the electron properties in the continuum limit, where the electron dynamics is governed by a Dirac equation defined on the curved graphene strip \cite{diracgraphene}. The curved geometry induces a pseudo-magnetic potential vector $\Gamma_\mu$, known as the spinorial connection. Then, we explore some features and effects of the spinorial connection, such as the geometric potential. 

We employ the intrinsic coupling of the electron to the curved surface, wherein the massless Dirac equation on the surface is given by \cite{diracgraphene}
\begin{equation}\label{d1}
    i\hbar \gamma^\mu D_\mu \psi = 0.
\end{equation}
In the Eq.(\ref{d1}), the gamma matrices on the surface $\gamma^\mu$ as defined as
\begin{equation}
\gamma\,^\mu = e^\mu_a \gamma^a,
\end{equation}
where $e^{a}_{\mu}$ are the dreinbeins. The spinorial covariant derivative $D_\mu$ is defined as
\begin{equation}
\label{cov}
D_\mu=\partial_\mu-\Gamma_\mu,
\end{equation}
where the spinorial connection $\Gamma_\mu$ is defined as
\begin{align}
\label{spinorconnection}
    \Gamma_\mu = \frac{1}{4} \omega^{a b}_{\mu}  \, \gamma_a \gamma_b.
\end{align}
Since the curved covariant derivative in Eq.(\ref{cov}) is similar to the minimal coupling between an electron and a magnetic vector $A_\mu$, the spinorial connection can be interpreted as a pseudo-magnetic potential vector induced by the curved geometry.
In the spinorial connection in Eq.(\ref{spinorconnection}), $\omega^{a b}_{\mu}$ are the connection 1-form and $\gamma_a$ are the flat Dirac matrices. In this work we use the following representation of the Dirac matrices $\gamma^0 = -i \sigma_3$, $\gamma^1 = -\sigma_ 2$, and $\gamma^2 = \sigma_1$ \cite{catenoid2}.

From Eq.(\ref{connection}), since $\omega^{1}_{2}=-(\partial_u \beta -1)d\theta$, the only non-vanishing component of the spinorial connection is given by
\begin{equation}
\label{spinconnection}
\Gamma_\theta = -\frac{i}{2}(\partial_u \beta -1)\sigma_3. 
\end{equation}
Thus, the pseudo-magnetic potential has only one component along the angular direction. Incidentally, the corresponding pseudo-magnetic field should point into the normal direction and it should be proportional to the Gaussian curvature \cite{diracgraphene}.

\begin{figure}
\begin{center}
\includegraphics[scale=0.25]{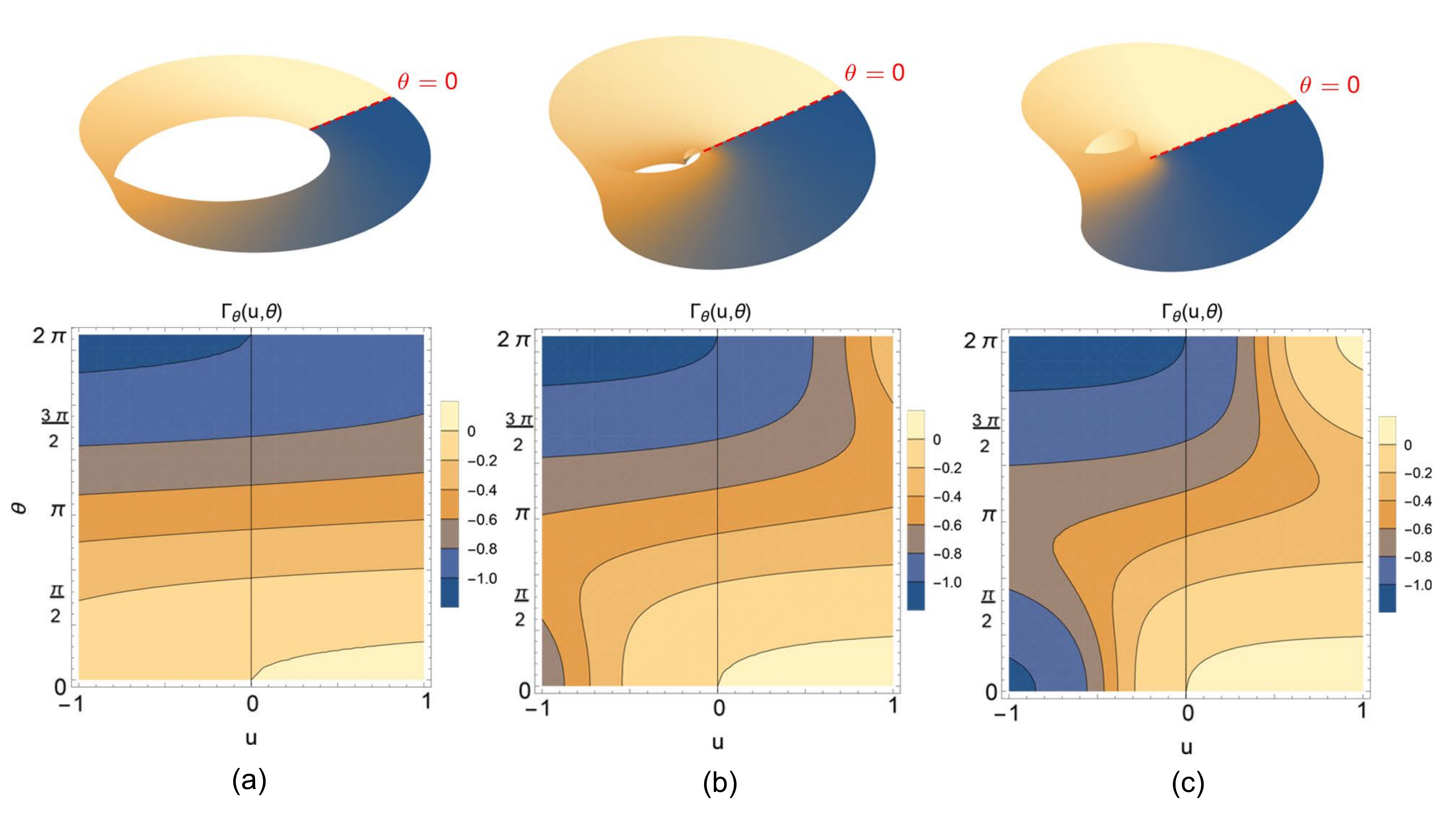}
\end{center}
\caption{Spinorial connection on the graphene M\"{o}bius strip with L/a = 0.375 (a), L/a = 1 (b) and L/a = 1.89 (c).}
\label{spinorconnectionfigure}
\end{figure}

The behaviour of the spinorial connection along the graphene M\"{o}bius strip is shown in the fig.(\ref{spinorconnectionfigure}). We adopt the same values for the ratio $L/a$ as done before, i.e.  $L/a=0.375$ (a), $L/a=1$ (b) and $L/a=1.89$ (c). Note that
the spinorial connection profile is rather different from the Gaussian and mean curvatures. Indeed, the connection is greater for inner points $(u<0)$ than for outer points $(u>0)$ of the strip. For the sake of comparison, the spinorial connection for a cone is constant $\Gamma_\theta = -\frac{i}{2}(\alpha -1)\sigma_3$ \cite{cone} and it vanishes for a single layer graphene, where $\alpha=1$.

\subsection{Hamiltonian}
Let us consider a stationary electronic states, i. e.,
\begin{equation}
\psi(\textbf{r},t)=\text{e}^{\frac{iEt}{\hbar}}\varphi(\textbf{r}). 
\end{equation}
Thus, the Dirac equation (\ref{d1}) reads
\begin{equation}
    H\varphi = E \varphi,
\end{equation}
where, $H$ is the stationary Hamiltonian of the relativistic electron at the graphene Möbius strip of form
\begin{align} \label{hamiltonian}
    H=-i\hbar v_F \bigg[\sigma^{1}\bigg(\partial_{u}-\frac{1}{2} \frac{(\partial_{u}\beta -1)}{\beta}\bigg)+\frac{\sigma^{2}}{\beta}\partial_{\theta}\bigg].
\end{align}
Using the hermiticity relations of Dirac matrices, $(\gamma^{i})^{\dagger}=\gamma^{0}\gamma^{i}\gamma^{0}$, it is possible to show that the Hamiltonian in eq.(\ref{hamiltonian}) is indeed Hermitian.
In matrix notation, the Dirac equation in terms of the spinors is
\begin{align}
   -i \begin{pmatrix}
   0 & \partial_u -\frac{i}{\beta}\partial_\theta- \frac{(\partial_u \beta -1)}{2\beta}  \\
   \partial_u +\frac{i}{\beta}\partial_\theta- \frac{(\partial_u \beta -1)}{2\beta} & 0
   \end{pmatrix} \begin{pmatrix} \varphi_1 \\ \varphi_2 \end{pmatrix}
   = k \begin{pmatrix} \varphi_1 \\ \varphi_2 \end{pmatrix},
   \label{hamiltonianmatrix}
\end{align}
where $\varphi_{1,2}(\mathbf{r})=\varphi_{1,2}(u,\theta)$ and $k=\frac{E}{\hbar v_F}$ is the wave-vector norm (momentum).

\begin{figure}
\begin{center}
\includegraphics[scale=0.25]{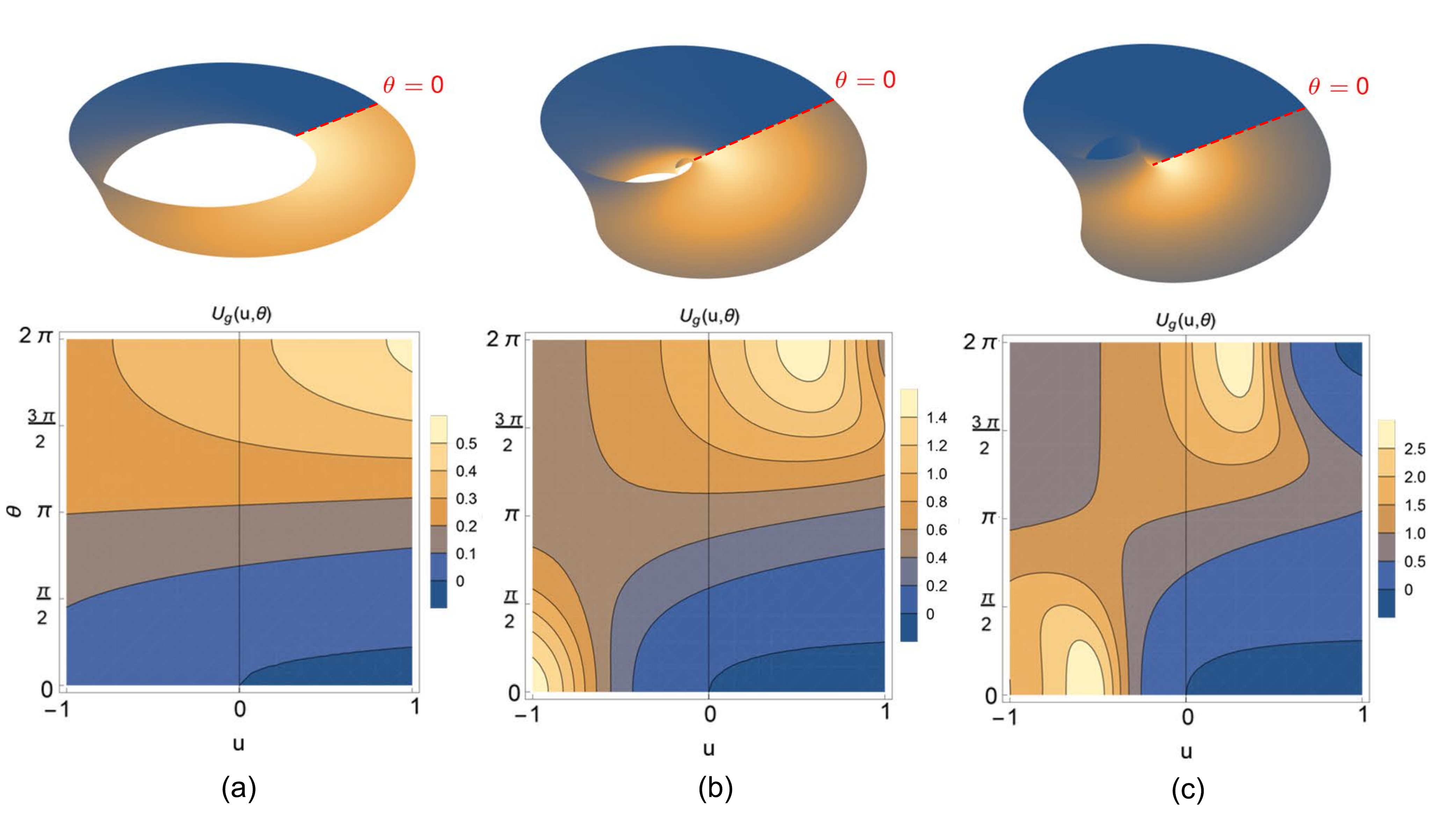}
\caption{Geometric potential $U_g$ on the M\"{o}bius strip with L/a = 0.375 (a), L/a = 1 (b) and L/a = 1.89 (c).}
\label{geometricpotential}
\end{center}
\end{figure}

It is worthwhile to mention that in the Hamiltonian Eq.(\ref{hamiltonianmatrix}) there is a geometric potential $U_g$ of the form
\begin{equation}
\label{connectionpotential}
U_g = -\frac{1}{2}\sigma^1 \frac{(\partial_u \beta -1)}{\beta}.
\end{equation}
The geometric potential $U_g$ has natural dimension of $L^{-1}$ and it steams from the coupling between the fermion and the spinorial connection. Note that for a conical surface, $U_g \approx \frac{(\alpha -1)}{u}$ \cite{cone}, and thus, the geometric potential vanishes for a flat single layer graphene sheet. We plotted the geometric potential for $L=1$ in the fig.(\ref{geometricpotential}), where  $L/a=0.375$ (a), $L/a=1$ (b) and $L/a=1.89$ (c). The geometric potential has regions with positive and negative values near each other. Moreover, the potential exhibits a sort of M\"{o}bius modified parity symmetry of the form $(u,\theta)\rightarrow (-u,2\pi -\theta)$.

Another noteworthy feature of the Hamiltonian in Eq.(\ref{hamiltonianmatrix}) is that it is no longer possible to write the wave function as $\varphi(u,\theta) = \text{e}^{i\ell \theta}\phi(u)$, as done for other surfaces such as the cone \cite{cone}, the helicoid \cite{watanabe}, catenoid \cite{catenoid,catenoid2} and the torus \cite{torus}. The reason for that steams from the lack of axial symmetry of the M\"{o}bius strip, which leads to a non-conservation of the angular momentum along the $z$ axis \cite{mobiusschrodinger}. As a result, no centrifugal term of form $\frac{l}{\beta}$ arises naturally in the Hamiltonian Eq.(\ref{hamiltonianmatrix}).


\section{Dirac fermion on wires}
\label{sec4}

In the last section we described how the electron dynamics is modified by the curved geometry of the graphene M\"{o}bius strip. In this section we explore the electronic properties of the Dirac fermion constrained to wires along the graphene M\"{o}bius strip. By doing so, the fermion dynamics is dependent not only on the wire curvature but on the geometric potential $U_g$ as well.

\subsection{Wires along the width}

Let us start with a wire directed along the strip width. In order to do it, we consider a given angle $\theta=\theta_0$
and vary the Hamiltonian in Eq.(\ref{hamiltonianmatrix}) only along the $u$ direction. Since the Hamiltonian is highly dependent on the angular coordinate $\theta$, we investigate how the electron properties change as we consider different direction on the strip.

Along the $u$ direction, we use $\beta = \beta(u,\theta_0)$ and $\partial_u \beta=\frac{\partial \beta(u,\theta)}{\partial u}\bigg\vert_{\theta = \theta_0}$. The Dirac equation Eq.(\ref{hamiltonianmatrix}) yields to
\begin{align}
    \left(  \partial_u- \frac{(1-\partial_u \beta)}{2\beta}  \right) \left(   \partial_u - \frac{(1-\partial_u \beta)}{2\beta} \right) \varphi_i = -k^2 \varphi_i,  \label{eq46} 
\end{align}
where $i$ can take on values $1$ or $2$. This equation, valid for both components of the spinor, can be rewritten as 
\begin{align}
    \partial_u^2 \varphi_i - \left( \frac{(1-\partial_u \beta)}{\beta} \right) \partial_u \varphi_i + \left[\left(\frac{(1-\partial_u \beta)}{2\beta}\right)^2 - \partial_u\left( \frac{(1-\partial_u \beta)}{2\beta} \right)  + k^2  \right] \varphi_i = 0 \label{eq52},
\end{align}
where the components of the spinor are decoupled and they satisfy the same Eq.(\ref{eq52}).

The Eq. (\ref{eq52}) can be further simplified by considering the change on the wave function \begin{align}\label{change1}
    \varphi_{i}(u)=e^{-\frac{1}{2}\int{\frac{(1-\partial_u \beta)}{\beta}}du}\chi_{i}(u),
\end{align}
where the new wave function $\chi_{i}(u)$ satisfies
\begin{align}\label{c3}
- \frac{d^2 \chi_i(u)}{du^2} = k^2 \chi_i(u) .
\end{align}
Accordingly, the exact wave function along the wire for $(u,\theta_0)$ is given by 
\begin{equation}
\varphi_i(u,\theta_0)= \sqrt{\beta(u,\theta_0)}e^{(-\frac{1}{2}\int{\frac{1}{\beta}}du)}\bigg[A\cos(ku)+B\sin(ku)\bigg].
\end{equation}
Since the factor in the brackets are the solution for a free particle constrained inside the range $-L\leq u \leq L$, the effect of the M\"{o}bius geometry on the electron constrained to a wire along the width is encoded in the 
geometric phase 
\begin{equation}
\Delta \phi = e^{-\frac{1}{2}\int{\frac{(1-\partial_u \beta)}{\beta}}du}.
\end{equation}
It is worthwhile to mention that, this geometric phase steams from the geometric potential $U_g$ in Eq.(\ref{connectionpotential}) which depends on the connection rather than on the surface curvature. In a cylindrical surface (nanotubes), $\partial_u \beta =0$ and the geometric phase is constant. On the other hand, for a flat plane (single layer graphene), $\beta^2=u^2$ and thus, the geometric phase vanishes identically. 

Imposing the boundary conditions 
\begin{align}\label{cond1}
\varphi_i(u=L,\theta_0)=\varphi_i(u=-L,\theta_0)=0,
\end{align}
one obtains 
\begin{align}
    A=0 \hspace{1cm} \text{and} \hspace{1cm} \sin(kL)=0,
\end{align}
for which the allowed energy spectrum is given by
\begin{align}
\label{energylevelwidth}
    E_n= n\frac{\pi}{L}\hbar v_F.
\end{align}
Note that the energy spectrum increases as the strip width $L$ decreases. Moreover, the energy levels grow linearly with $n$ and $\hbar v_F$, as expected from the Dirac equation. In the Ref.(\cite{mobiusschrodinger}), the authors found the energy spectrum for a non-relativistic electron in the M\"{o}bius band is proportional to $n^2$, an expected result steaming from the Schr\"{o}dinger equation. In addition, the energy levels found in Eq.(\ref{energylevelwidth}) is similar to the spectrum in a cylindrical ring. Indeed, since the geometric potential effects are encoded into the phase, the spectrum is the same for a flat surface. For a ring with the same width as one studied in Ref.(\cite{ABring}), i.e., for $2L=150 nm$, a typical electron with Fermi velocity $v_F =c/300$ has a ground state energy about $E_0 \approx 2.75\times 10^{-2} eV$.

Therefore, the wave function along the wire is given by
\begin{align}
    \varphi_1(u,\theta_0)=C\sqrt{\beta(u,\theta_0)}e^{(-\frac{1}{2}\int{\frac{1}{\beta}}du)}\sin\bigg(\frac{n\pi u}{L}\bigg).
\end{align}

\begin{figure}[ht!]\centering
           \includegraphics[width=8cm,height=5cm]{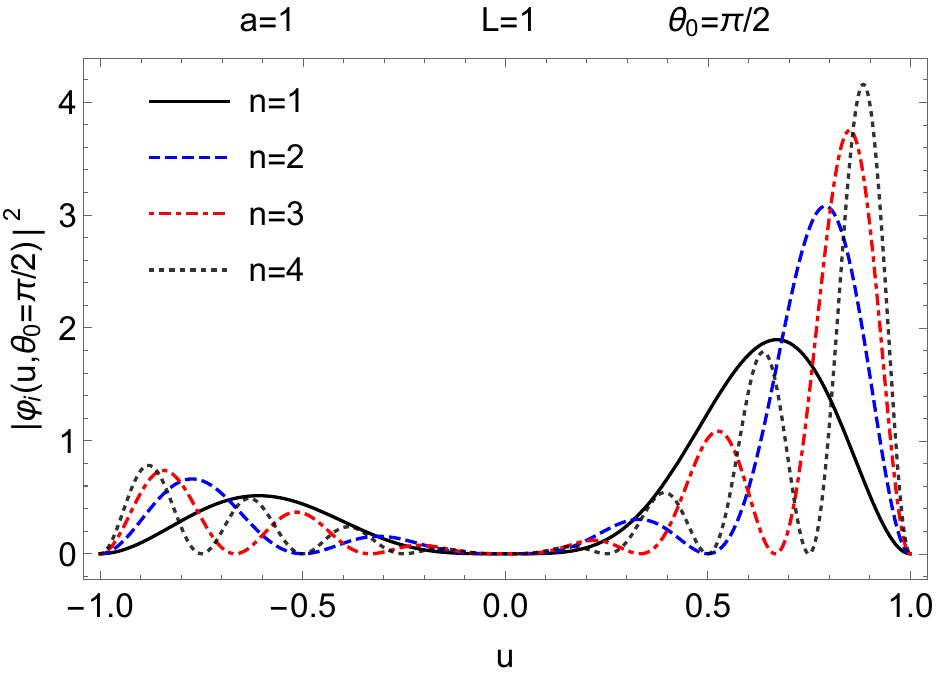}
           \includegraphics[width=8cm,height=5cm]{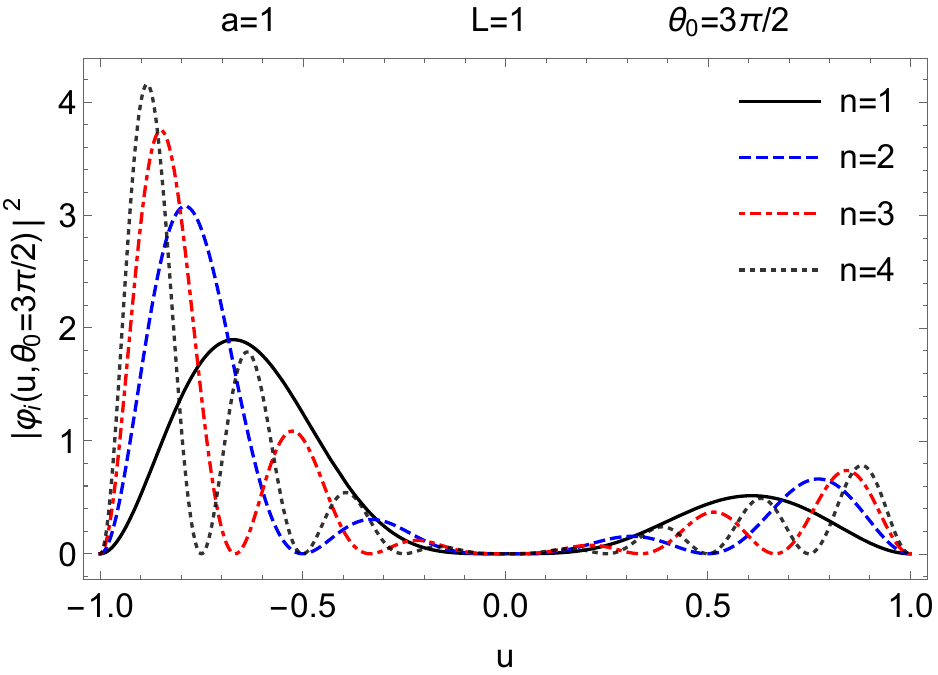}
           \includegraphics[width=8cm,height=5cm]{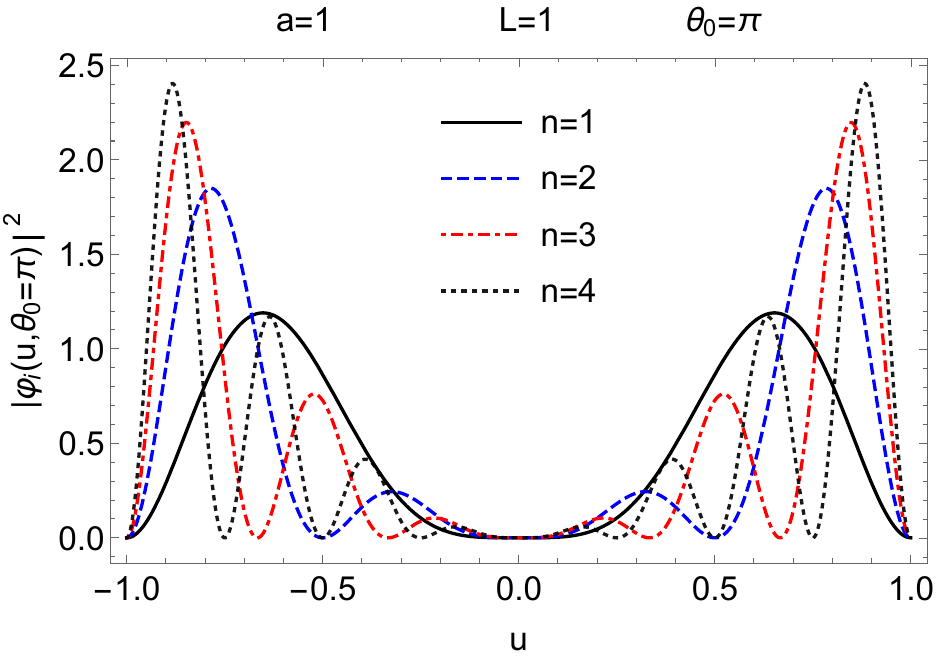}
           \includegraphics[width=8cm,height=5cm]{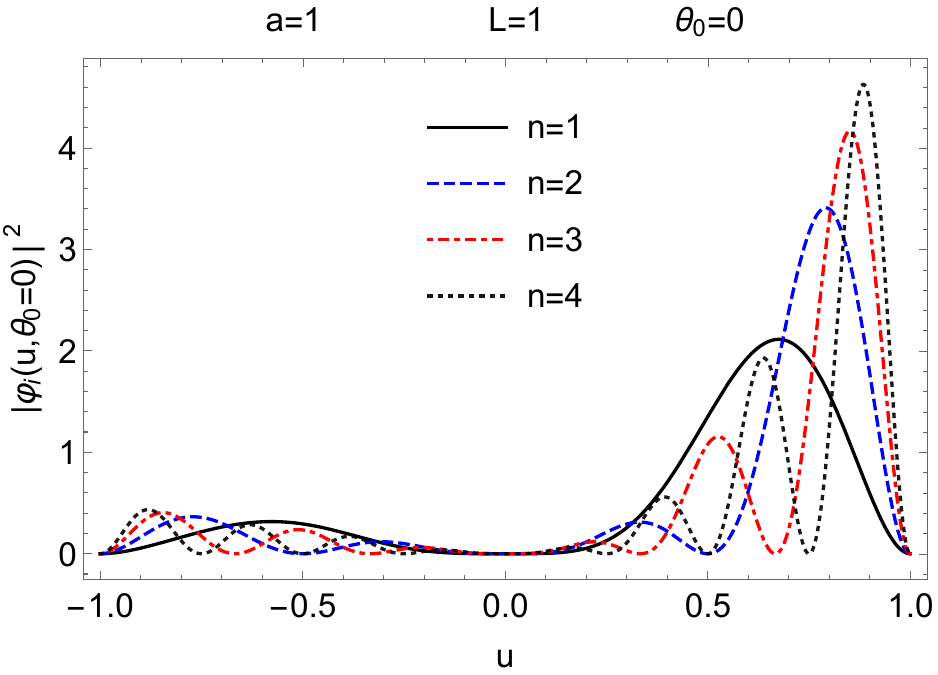}
 \caption{Density of states along the width for $\theta_0$ fixed. It shows the effect of the geometric phase concentrating the wave function at the outer region of the strip, leading to the formation of edge states.}
 \label{fig:u}
\end{figure}

The effects of the geometric phase on the electron is shown in the Fig.(\ref{fig:u}), where we plotted the probability density $\varphi(u,\theta_0)\bar{\varphi}(u,\theta_0)$ for the four first energy levels. It is worthwhile to mention that the region where the fermion is localized on the wire depends on which angle $\theta_0$ the wire is on the strip. Indeed, for $\theta_0 =\frac{\pi}{2}$, the wave function is concentrated at the outer region of the strip, whereas for $\theta_0 =\frac{3\pi}{2}$, the fermion is more localized at the inner region. Thus, the angle $\theta_0$ can be understood as a parameter to tune the region where the electron is more concentrated. Moreover, note that the probability density does not possesses  parity symmetry for $\theta_0 =\{0, \pi/2 , 3\pi/2 \}$. 

Another noteworthy feature shown in the fig.(\ref{fig:u}) is that the geometric phase tends to a damping of the amplitude of the Dirac fermion. In fact, by comparing the graphics of the wave function and the geometric potential, we can see that the fermion is more concentrated in the regions where the potential is less strong.

\begin{figure}[ht!]\centering
           \includegraphics[scale=0.24]{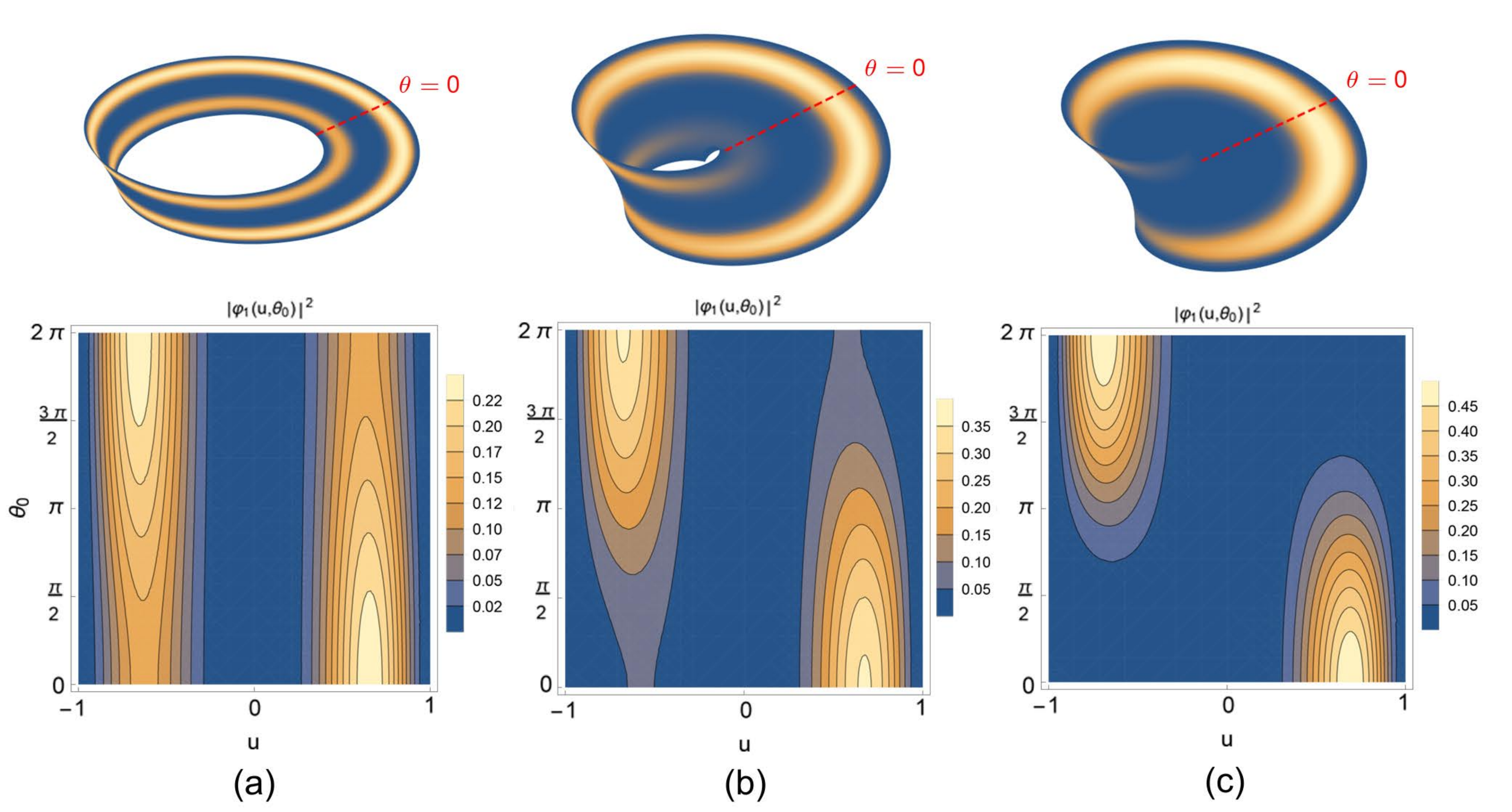}
 \caption{Density of states along the M\"{o}bius surface for n=1 with L/a = 0.375 (a), L/a = 1 (b) and L/a = 1.89 (c). The electron is more concentrated around the edges of the strip.}
 \label{densityofstatesu}
\end{figure}

\begin{figure}[ht!]\centering
           \includegraphics[scale=0.24]{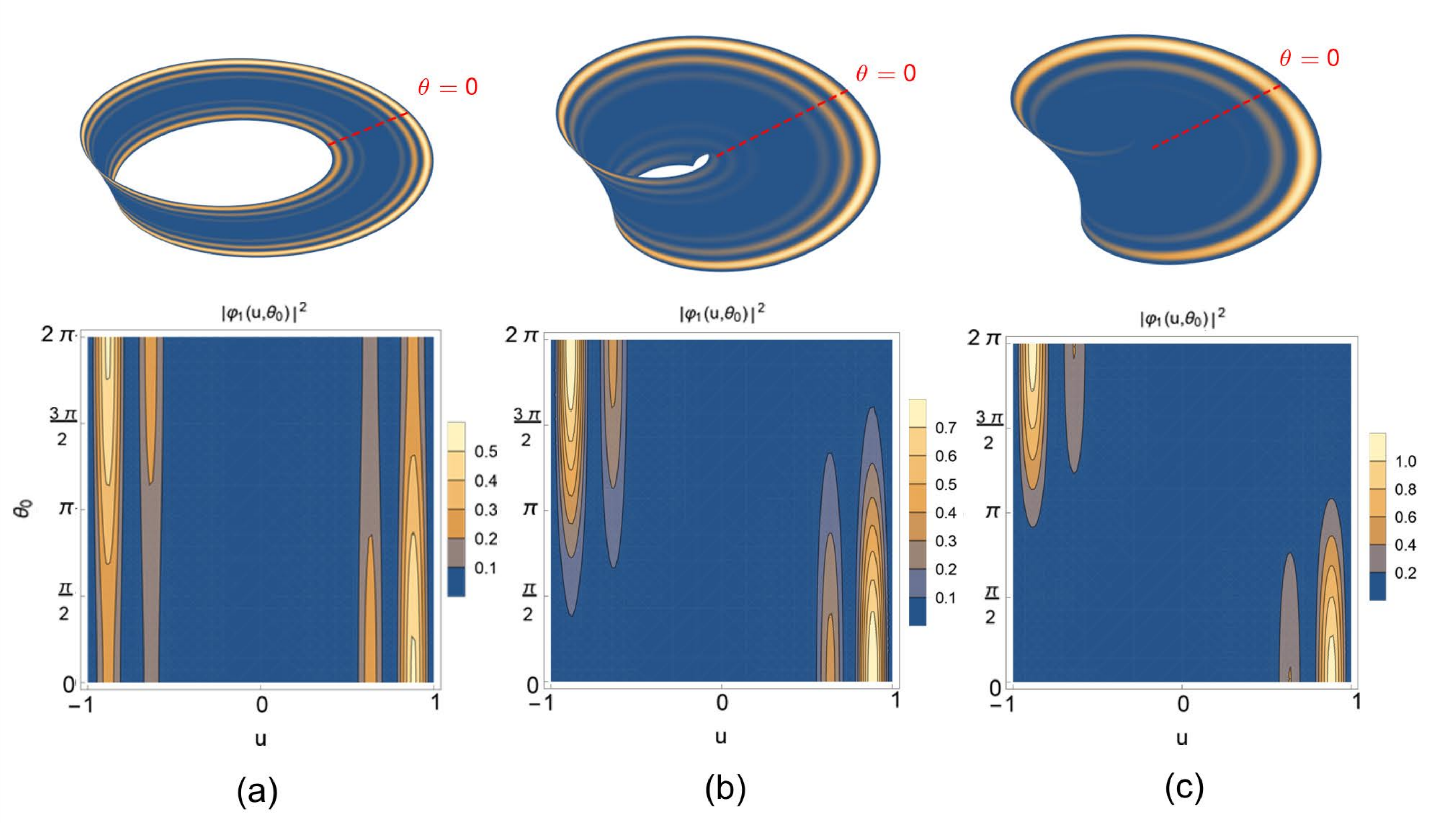}
 \caption{Density of states along the M\"{o}bius surface for n=4 with L/a = 0.375 (a), L/a = 1 (b) and L/a = 1.89 (c). Note the formation of more rings near the edges of the strip.}
 \label{densityofstatesu2}
\end{figure}

In fig.(\ref{densityofstatesu}), we plot the density of states on the M\"{o}bius strip for $n=1$. We vary the ratio $a/L$ for $L/a=0.375$ (a), $L/a=1$ (b) and $L/a=1.89$, where we can see the formation of edge states. The presence of robust edge states has already been pointed out in Ref.\cite{mobiustopologicalinsulator}, where the topological insulator behaviour of the M\"{o}bius strip was investigated. Moreover, note that the ground state $(n=1)$ in fig.(\ref{densityofstatesu}) exhibits only one ring state whereas the excited state for $(n=4)$ in fig.(\ref{densityofstatesu2}) shows more ring states.

\subsection{Wire along the strip length}

Let us now consider the Dirac fermion constrained in a wire for a fixed $u=u_0$. For this case, we have no momentum along $u$, so that $\partial_u \psi = 0$. Also, $\beta = \beta(u_0,\theta)$ and $\partial_u \beta = \frac{\partial \beta(u,\theta) }{\partial u} |_{u = u_0}$. Thus, by applying  the Hamiltonian (\ref{hamiltonian}) to the wavefunction $\psi$, we obtain
\begin{align} 
\label{hamiltonianalonglength}
    H\psi=-i\hbar v_F \bigg[-\frac{1}{2} \frac{(\partial_{u}\beta -1)}{\beta}\sigma^{1}+\frac{\sigma^{2}}{\beta}\partial_{\theta}\bigg]\psi.
\end{align} 
By making the change in the wave function
\begin{align}
\psi(\theta,u_0) = e^{\Gamma_\mu x^\mu}  \psi_0 = e^{-\frac{i}{2} \frac{(\partial_{u}\beta -1)}{\beta} \sigma^3 } \psi_{0}(\theta,u_0),
\end{align} 
we can simplify the equation (\ref{hamiltonianalonglength}) to
\begin{align}
H\psi_0 =-i\hbar v_F \frac{\sigma^{2}}{\beta}\partial_{\theta}\psi_0.
\end{align}
Thus, the Dirac equation $H\psi_0 = E\psi_0$ for the spinor $\psi_0$ whose components are of form
\begin{align}
\psi_0 = \begin{pmatrix}
\Phi_1(\theta) \\ \Phi_2(\theta) 
\end{pmatrix},
\end{align}
yields to to the system
\begin{align} 
\label{angulardiracequation}
   -i \begin{pmatrix}
   0 &  -\frac{i}{\beta}\partial_\theta  \\
   \frac{i}{\beta}\partial_\theta & 0
   \end{pmatrix} \begin{pmatrix} \Phi_1 \\ \Phi_2 \end{pmatrix}
   = k \begin{pmatrix} \Phi_1 \\ \Phi_2 \end{pmatrix},
\end{align}
where $k = \frac{E}{\hbar v_F}$. By decoupling the system in eq.(\ref{angulardiracequation}) leads us to a differential equations valid for both components of the spinor given by
\begin{align} \label{ODEtheta}
-\frac{1}{\beta} \frac{d}{d\theta} \bigg( \frac{1}{\beta} \frac{d\Phi_i}{d\theta}  \bigg) = k^2 \Phi_i(\theta).
\end{align}
The Eq.(\ref{ODEtheta}) can be further simplified by considering the change of coordinate
\begin{align}
    v(\theta) = \int_0^\theta \beta(u_0,\theta') d\theta',
\end{align}
where $dv=\beta(u_0 , \theta)d\theta$ is the infinitesimal arc length.
Accordingly, the decoupled Dirac equation Eq.(\ref{ODEtheta}) reads
\begin{align}\label{eq:edo1}
  \frac{d ^2\Phi_i(v)}{dv^2}+k^2\Phi_i(v) = 0,
\end{align}
whose exact solution for each component of the spinor $\psi$ along the strip length are given by
\begin{eqnarray}
\label{solutiontheta1}
\varphi_1(\theta) & = & e^{iW(\theta)} \bigg[ A \cos \bigg(k \int_{0}^{\theta} \beta(u_0,\theta') d\theta'\bigg) + B \sin \bigg(k \int_{0}^{\theta} \beta(u_0,\theta') d\theta'\bigg)  \bigg]\nonumber\\
\varphi_2(\theta) & = & e^{-iW(\theta)} \bigg[ C \cos\bigg(k \int_{0}^{\theta} \beta(u_0,\theta') d\theta'\bigg) + D \sin \bigg(k \int_{0}^{\theta} \beta(u_0,\theta') d\theta'\bigg)   \bigg], 
\end{eqnarray}
where the geometric phase $W(\theta)$ along the angular wire is given by
\begin{equation}
W(\theta) = -\frac{1}{2} \frac{(\partial_{u}\beta -1)}{\beta}.
\label{angulargeometricphase}
\end{equation}

It is worthwhile to mention that, despite the geometric phase $e^{iW(\theta)}$ modifies the wave function, 
the probability distribution $\varphi\bar{\varphi}$ is independent of $W$. This is key feature of the geometric phases \cite{cone}, and similar property is also shared with the Aharonov-Bohm phase \cite{ABring}.

Another noteworthy result is related to the period of the wave function. Indeed, by eq.(\ref{solutiontheta1}) the period
strongly depends on the metric function $\beta(u_0 , \theta)$. Thus, 
let us now investigate the effects of the geometry of the angular wires on the electronic states.


\subsubsection{Central ring}

At the center of the strip, i.e., for $u=0$ the wire forms a closed ring. The angular metric factor $\beta$ takes the form
\begin{align}
    \beta(u_0 = 0, \theta) = a,
\end{align}
and thus, the metric on this ring is independent of $\theta$. As a result, the geometric phase has the form
\begin{equation}
W(\theta)=\frac{1}{2}\left(1-\cos\left(\frac{\theta}{2}\right)\right),
\end{equation}
and the first spinor component wave function is given by
\begin{equation}
\label{ringwavefunction}
\varphi_1(\theta) = e^{i\frac{1}{2}\left(1-\cos\left(\frac{\theta}{2}\right)\right)} \bigg[ A \cos ka\theta + B \sin ka\theta  \bigg].
\end{equation}
The second spinor component wave function can be obtained from Eq.(\ref{ringwavefunction}) by making $W\rightarrow -W$.
Interestingly, although the wire for $u=0$ forms a circular ring, the M\"{o}bius strip still induces an anisotropic geometric phase. This result shows the difference between an usual ring and one constrained on the M\"{o}bius surface.
\begin{figure}[ht!]
    \centering
    \includegraphics[scale=0.5	]{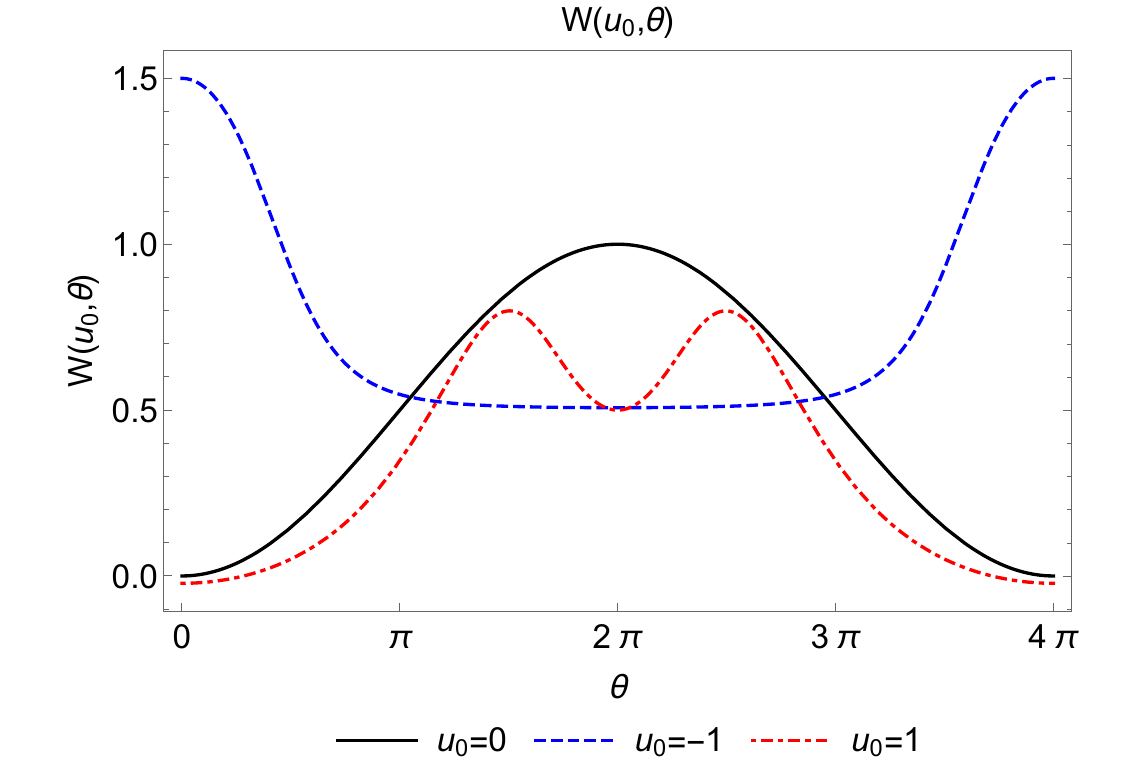}
    \caption{Geometric phase along the $\theta$ direction for $u_0$ = $0, \pm1$ showing its $4\pi$ periodic behaviour.}
    \label{geometricphasetheta}
\end{figure}

Even though the geometric phase $W(\theta)$ is a $4\pi$ periodic function, as shown in the Fig.(\ref{geometricphasetheta}), the period of the wave function is determined by the trigonometric function inside the brackets. In fact, by considering the periodic boundary conditions 
\begin{equation}
\label{boundaryconditioncentralring}
\varphi_1(0,\theta= 0)=\varphi_1(0,\theta= 2\pi)=0,
\end{equation}
the wave function has the form
\begin{align}
    \varphi_1 (\theta) = A e^{iW(\theta)} \sin \big(a k \theta \big). \label{eq:func}
\end{align}
Since $e^{iW(\theta)}$ lies at the unit circle in the complex plane, this factor never vanishes. Thus, the wave function period is determined by the $\sin ka\theta$ function.

Using the boundary condition on the central ring in Eq.(\ref{boundaryconditioncentralring}), the momentum is given by
\begin{equation}
k_n=\frac{n}{2a},
\end{equation}
and thus, the wave function has the form
\begin{equation}
    \varphi_1 (\theta) = A e^{iW(\theta)} \sin \left(n \frac{\theta}{2}\right),
\end{equation}
where $n$ is a integer number. It is worthwhile to mention that the ground state $(n=1)$ has a period of $4\pi$, whereas the first excited state $(n=2)$ is a $2\pi$ periodic function. For the $n-$th state, the period is $T_n = 4\pi/n$, and hence for $n$ odd, the period is a non-integer multiple of $4\pi$. This feature results from the M\"{o}bius strip geometry and similar results were found for a non-relativistic electron \cite{mobiusschrodinger}.

\begin{figure}[ht!]
    \centering
    \includegraphics[scale=0.5]{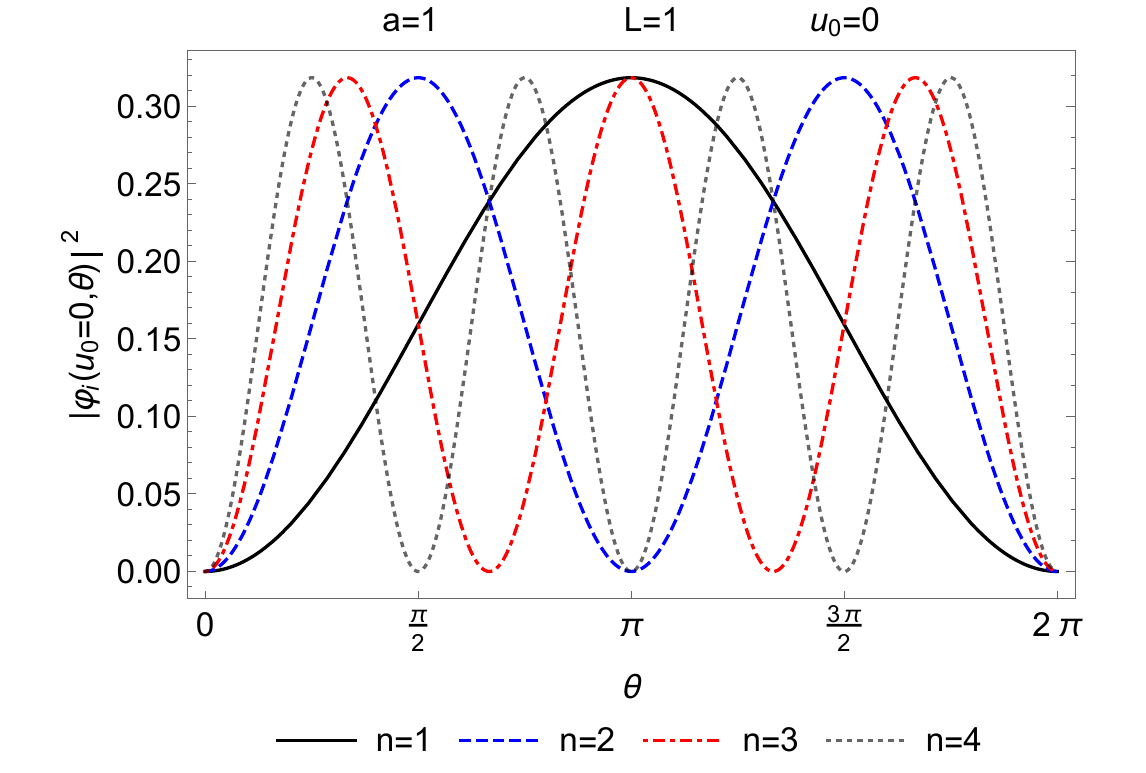}
    \caption{Density of states at the central ring ($u=0$) for the initial four energy levels. The wave function shows a free particle behaviour, since in this case the geometric phase doesn't affect the density probability.}
    \label{wavefunctioncentralring}
\end{figure}

In the fig.(\ref{wavefunctioncentralring}) we plotted the probability density $\psi\bar{\psi}$ for the first electronic states. The figure shows that the ground state $(n=1)$ is centred at $\theta=\pi$, whereas the first excited state $(n=2)$ has two peaks 
symmetrically displaced from $\theta=\pi$. As $n$ increases, the number of peaks increases as well. Besides, for odd $n$ the probability density has a peak at $\theta=\pi$, whereas for even $n$ the probability vanishes at this point.

Accordingly, the allowed energies $E_n$ are given by
\begin{align}\label{eq:espectro}
    E_n= \frac{n}{2} \frac{\hbar v_F}{a}, 
\end{align}
Note that the energy decreases as the inner radius $a$ increases, as expected for an usual ring. Furthermore, it is important to mention that the spectrum is the same for both components of the spinor. However, the nontrivial topology leads to an energy spectrum which is the half-integer multiple of the ground state, shown in Eq.(\ref{eq:espectro}). A similar result for a non-relativistic electron was found in the Ref.(\cite{mobiusschrodinger}). For a ring with the same 
inner radius as one studied in the Ref.(\cite{ABring}), i.e., for $a=200 nm$, the energy of the ground state is $E_0 \approx 1,64\times 10^{-3} eV$, which is lower than the ground state energy for a wire along the width. It is worthwhile to mention that the values $a=200 nm$ and $L=75 nm$ leads to the ratio $L/a=0,375$ and thus, they can be used to form stable M\"{o}bius strip \cite{mobiusstrip1}.

\subsubsection{Edge wires}


For wires at the edge of the M\"{o}bius band, i.e., for $u=\pm 1$, the wires do not form closed rings. 
The arc length variable $v$ for the outer edge is given by
\begin{align}
    v(\theta)\vert_{u=1}=\int_0^\theta \bigg[\frac{1}{4}+(1+\cos (\theta'/2))^2\bigg]^{1/2} d\theta',
\end{align}
whereas for the inner edge has the form
\begin{align}
    v(\theta)\vert_{u=-1}=\int_0^\theta\bigg[\frac{1}{4}+(1-\cos(\theta'/2))^2\bigg]^{1/2} d\theta',
\end{align}
where we assume that $-1\leq u\leq 1$. Since the $\beta(\pm 1,\theta)$ is periodic of with period $4\pi$, then $v(\theta)$ is also a $4\pi$ periodic function. Therefore, the wave function has the form
\begin{eqnarray}
\varphi_1(\theta) & = & e^{iW(\theta)} \bigg[ A \cos \bigg(k v(\theta)\bigg) + B \sin \bigg(k v(\theta)\bigg)  \bigg].
\end{eqnarray}
Since $v(0)=0$, using the boundary conditions, the wave function is given by
\begin{equation}
\varphi_1(\theta) = A e^{iW(\theta)}\sin [k_n v(\theta)],
\end{equation}
where $k_n$ satisfies the condition
\begin{align}
     k_n \int_0^{4\pi} \beta(u_0,\theta) d\theta  = n \pi. \label{eq:espectro2}
\end{align}

\begin{figure}[ht!] 
        \centering
           \includegraphics[width=8cm,height=5cm]{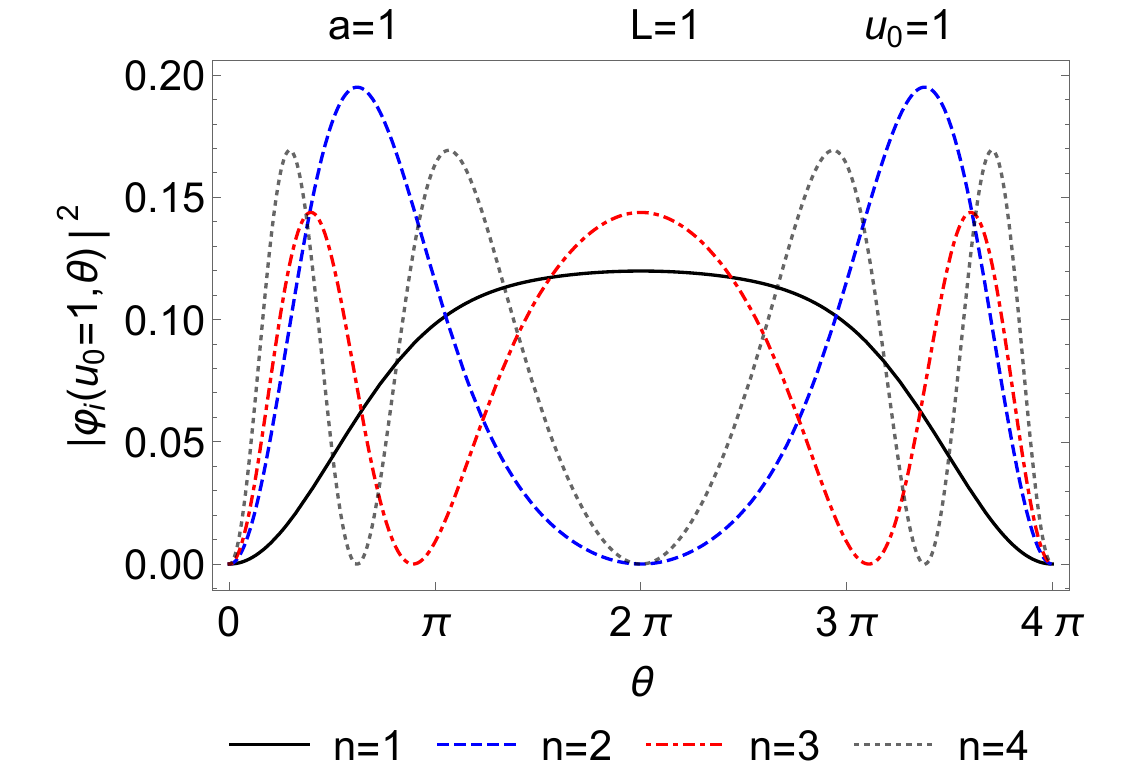}
           \includegraphics[width=8cm,height=5cm]{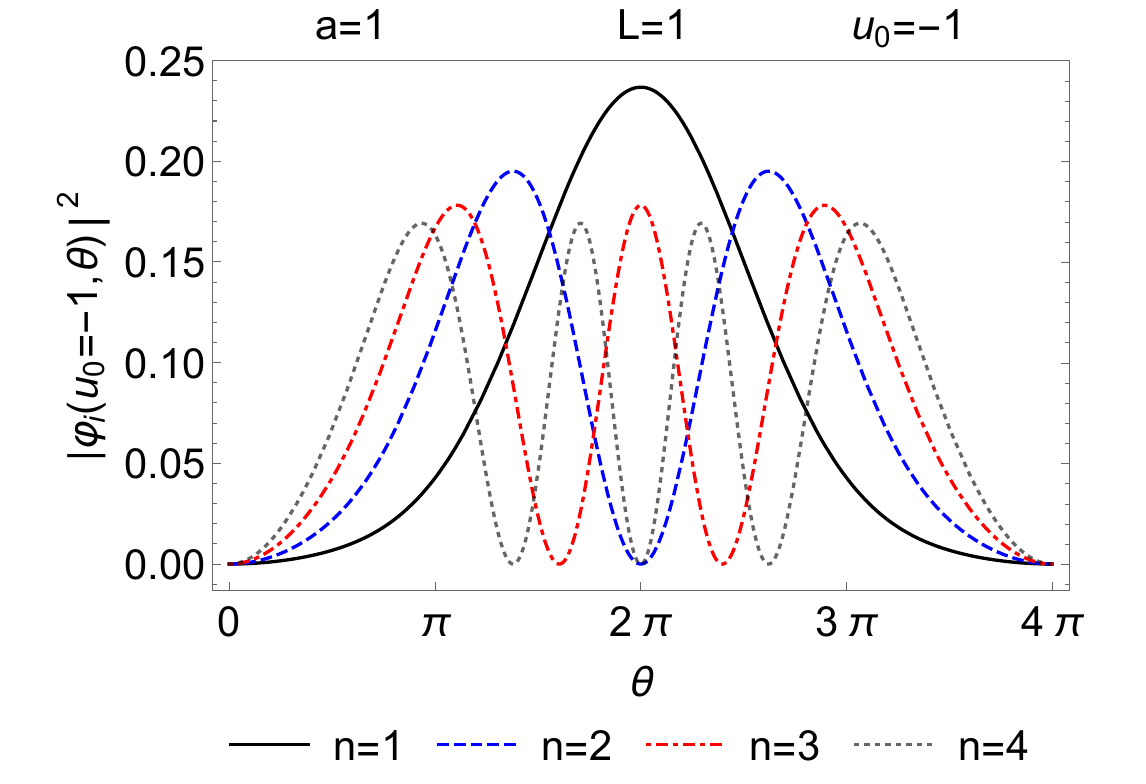}
       \vspace{-0.5cm}
\caption{Density of states at the edge wires ($u=\pm 1$) for the initial four energy levels.}
 \label{fig:func4}
 \end{figure}

We plot the probability densities in Fig. \ref{fig:func4}. Note that the probability associated with the
ground state is concentrated at $\theta=2\pi$ for $u=-1$ and $u=1$. Moreover, the first excited state has two peaks 
shifted from the point $2\pi$. As $n$ increases, the number of peaks also increases. Besides, the pattern of the wave functions for even $n$ differs from those for odd $n$.

\begin{figure}[ht!]\centering
           \includegraphics[scale=0.24]{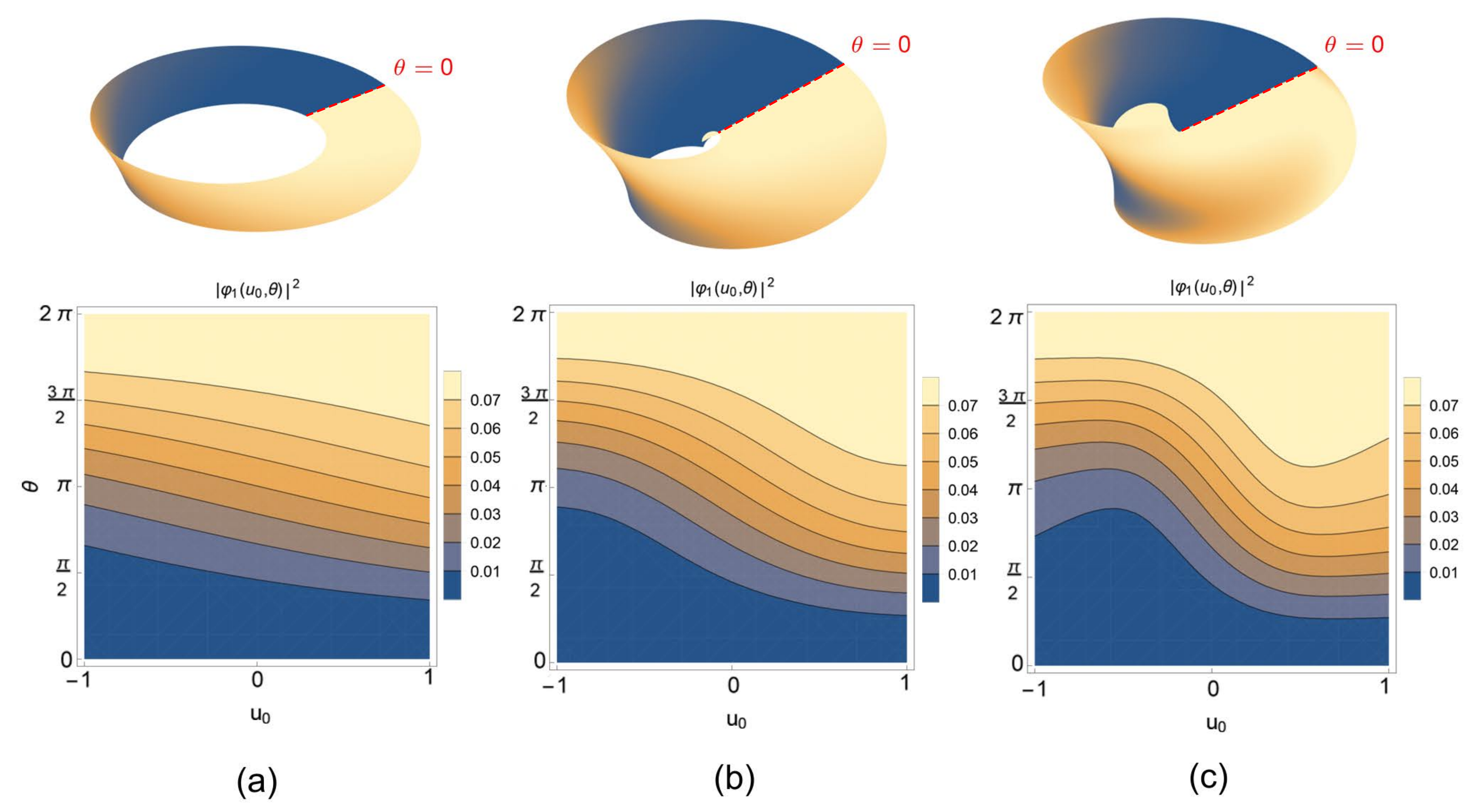}
 \caption{Density of states along the M\"{o}bius strip for n=1 with L/a = 0.375 (a), L/a = 1 (b) and L/a = 1.89 (c).}
 \label{densityofstatest}
\end{figure}

\begin{figure}[ht!]\centering
           \includegraphics[scale=0.24]{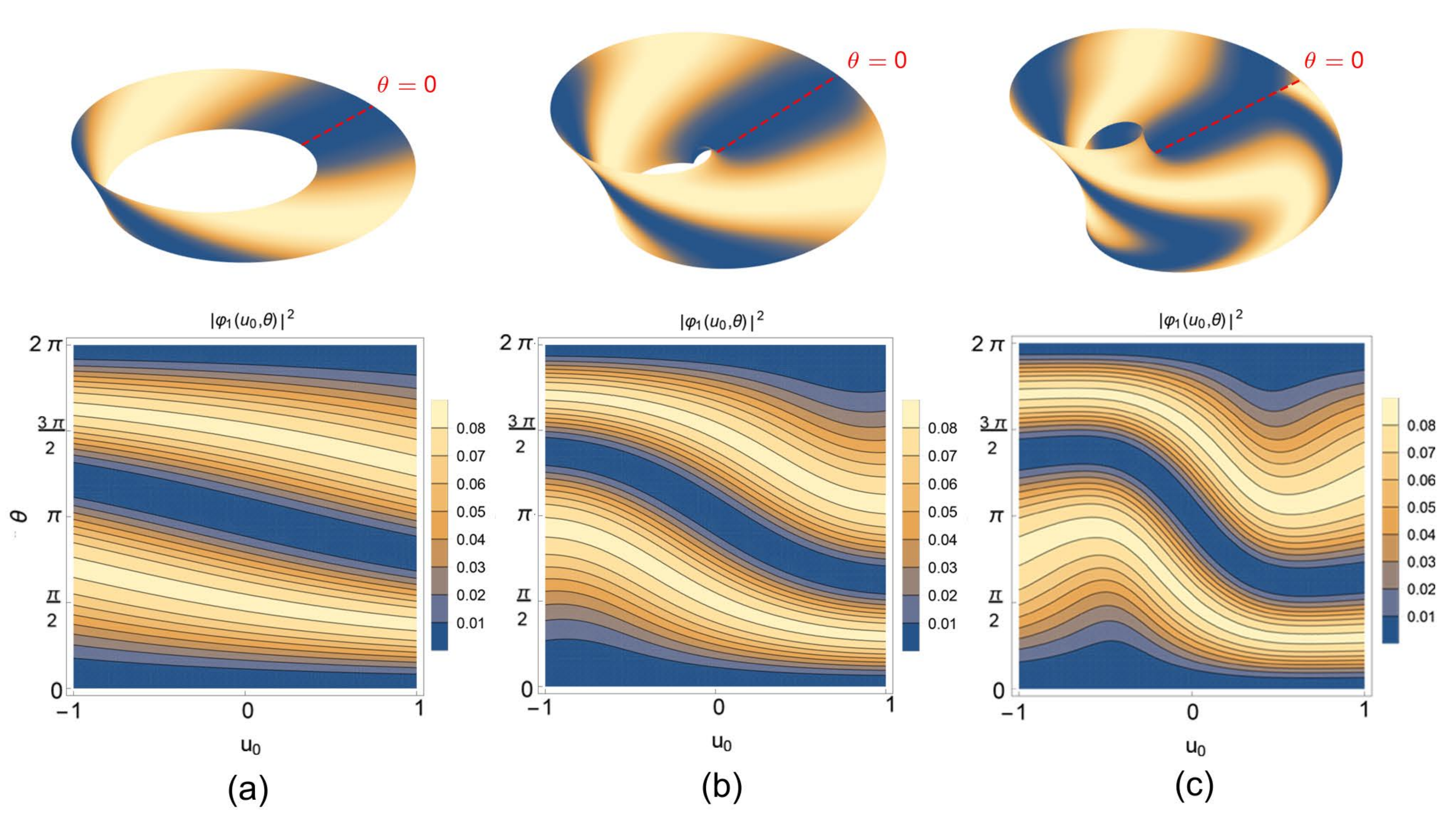}
 \caption{Density of states along the M\"{o}bius strip for n=4 with L/a = 0.375 (a), L/a = 1 (b) and L/a = 1.89 (c).}
 \label{densityofstatest2}
\end{figure}

The fig.(\ref{densityofstatest}) shows how the density of states are distributed on the M\"{o}bius strip for $n=1$. Note that the ground state (upper graphics) resembles the geometric potential profile shown in fig.(\ref{geometricpotential}). For $n=4$, shown in the fig.(\ref{densityofstatest2}), note that the density of states are rather dependent on the ratio $a/L$.

\section{Discussion and perspectives} 
\label{sec5}

In this work we found exact solutions for an electron constrained to wires along a graphene M\"{o}bius strip. Such a surface can be formed by performing a twist on one graphene ribbon end and connecting it to the another end. By considering the electron effectively described as a massless Dirac fermion, the M\"{o}bius strip curved geometry
provides a geometric potential depending on the geometric connection instead of the curvature. The geometry of the wires depends not only on the wire metric but on the M\"{o}bius geometry and symmetries, as well.

For wires along the strip width, i.e., for a given $\theta_0$, the effects of the curved geometry are encoded into a geometric phase which steams from the geometric potential. As a result, exact solutions for the Dirac equation were found
which shows the formation of edge states. As we increase the energy, additional edge states forming rings appear. Moreover, the electronic probability density is highly dependent on the direction $\theta_0$ chosen for the wire. In fact, for some $\theta_0$ the electron is more concentrated 
near the inner edge, whereas for other values of $\theta_0$ the wave function is localized towards the outer edge of the band. This behaviour agrees with the lack of parity symmetry of the M\"{o}bius band, i.e., $\beta(-u,\theta_0)\neq \beta(u,\theta_0)$.

On the other hand, the wires along the strip length have a rather different geometry to those along the width. For $u=0$, the wire forms a ring of radius $a$, whereas for $u=\pm L$ (strip edges) the wire is open in the interval $0\leq \theta \leq 2\pi$. Due to the M\"{o}bius strip geometry, the wires for $u_0 \neq 0$ are actually close if we consider the period $4\pi$. Accordingly, we expect that the wave function to be a periodic function of period $4\pi$. It turns out that, for the central ring at $u=0$, the ground state is a $4\pi$ periodic function whose probability density is localized around $\theta=\pi$. For $n$ odd, the wave function has a period that is a non-integer multiple of $2\pi$. Similarly, the energy levels are half-integer multiple of the quantum $\hbar v_F/a$. These non-integer features of both the electronic states and the spectrum are the result of the non-trivial geometry of the M\"{o}bius band. For a ring with the same 
inner radius as the cylindrical ring studied in the Ref.(\cite{ABring}), i.e., for $a=200 nm$, the energy of the ground state is $E_0 \approx 1,64\times 10^{-3} eV$. Since the ratio $L/a=0.375$ is smaller than the critical ratio, the non-trivial topology of the M\"{o}bius strip could be used to generate a geometric Aharonov-Bohm effect due to pseudo-magnetic field.

For the wires along the edges of the M\"{o}bius strip, the ground states $n=1$ exhibit a localized probability density
around $\theta=2\pi$. Once again, the break of the parity symmetry yields to a different profile between the ground state at the inner edge $(u_0 = -L)$ and at the outer edge $(u_0 = L)$. The electronic states for the outer edge are more concentrate than those at the inner edge.

A remarkable result we found is the role played by the geometric phase on the electronic states. For the wires along the width, the Dirac equation can be simplified by considering a geometric phase depending on the geometric potential. It turns out that the energy levels and the period of the electronic states are determined as if the strip were flat. A similar result was found numerically by Ref.(\cite{mobiusschrodinger}), though the authors considered a non-relativistic electron. On the other hand, the geometric phase provides a damping of the wave function, what leads to the parity breaking profile discussed above. Along the angular direction, the geometric phase is given by the geometric (spinor) connection. Likewise the well-known Aharonov-Bohm phase, the geometric phase does not alter the density of states for a single electron. Hence, the period of the electronic states and the energy levels are determined by the angular metric function along the wire, i.e., $\beta(u_0 , \theta)$. This shows that the electronic properties on wires along the M\"{o}bius strip are different of those in usual circular wires, for the wires inherit the
anisotropic M\"{o}bius geometry.

This work suggests as a perspective the investigation of the effects of the geometric phase on the interference pattern 
between electrons. Indeed, Aharanov-Bohm effects on graphene  gated cylindrical nanorings of width $L=75 nm$ and inner radius $a=200nm$ $(L/a=0.375)$ were  investigated \cite{ABring}. By considering a M\"{o}bius ring with the same size, similar effects could be observed due to the geometric phase. Moreover, the inclusion of external fields might lead to an additional parameter to control the density of states. In addition, the interaction of the confined electron to the external electric or magnetic fields should leave a peculiar signature which could be used to characterize the material. 
In fact, as proposed in Ref.\cite{diracgraphene} transmission electron microscopy (TEM) or scanning tunneling microscopy (STM) can be used to probe the morphology and relate it to the density of states. For the M\"{o}bius graphene strip, these microscopic techniques could be used to determine the ratio $L/a$ of the samples.



\section*{Acknowledgements}
\hspace{0.5cm}The authors thank the Conselho Nacional de Desenvolvimento Cient\'{\i}fico e Tecnol\'{o}gico (CNPq), grants   n$\textsuperscript{\underline{\scriptsize o}}$ 312356/2017-0 (JEGS),  n$\textsuperscript{\underline{\scriptsize o}}$ 309553/2021-0 (CASA) for financial support.

\end{document}